\documentclass[times,twocolumn,final]{elsarticle}

\usepackage{medima}
\usepackage{framed,multirow}

\usepackage{amssymb}
\usepackage{amsmath}
\usepackage{latexsym}

\usepackage{url}
\usepackage{xcolor}
\usepackage{booktabs}
\usepackage{hyperref}

\usepackage{graphicx}
\usepackage{booktabs}

\definecolor{newcolor}{rgb}{.8,.349,.1}

\journal{Medical Image Analysis}

\begin{document}

\verso{X. Guo, B. Zhou, D. Pigg, \textit{et~al.}}

\begin{frontmatter}

\title{Unsupervised inter-frame motion correction for whole-body dynamic PET using convolutional long short-term memory in a convolutional neural network} 

\author[1]{Xueqi Guo}
\author[1]{Bo Zhou}
\author[2]{David Pigg}
\author[2]{Bruce Spottiswoode}
\author[2]{Michael E. Casey}
\author[1,3]{Chi Liu\corref{cor1}}
\author[1,3]{Nicha C. Dvornek\corref{cor1}}
\cortext[cor1]{Corresponding author. Email: nicha.dvornek@yale.edu, chi.liu@yale.edu}

\address[1]{Department of Biomedical Engineering, Yale University, New Haven, CT 06511, USA}
\address[2]{Siemens Medical Solutions USA, Inc., Knoxville, TN, 37932, USA}
\address[3]{Department of Radiology and Biomedical Imaging, Yale University, New Haven, CT 06511, USA}

\received{DD MM 2021}

\begin{abstract}
Subject motion in whole-body dynamic PET introduces inter-frame mismatch and seriously impacts parametric imaging. Traditional non-rigid registration methods are generally computationally intense and time-consuming. Deep learning approaches are promising in achieving high accuracy with fast speed, but have yet been investigated with consideration for tracer distribution changes or in the whole-body scope. In this work, we developed an unsupervised automatic deep learning-based framework to correct inter-frame body motion. The motion estimation network is a convolutional neural network with a combined convolutional long short-term memory layer, fully utilizing dynamic temporal features and spatial information. Our dataset contains 27 subjects each under a 90-min FDG whole-body dynamic PET scan. With 9-fold cross-validation, compared with both traditional and deep learning baselines, we demonstrated that the proposed network obtained superior performance in enhanced qualitative and quantitative spatial alignment between parametric $K_{i}$ and $V_{b}$ images and in significantly reduced parametric fitting error. We also showed the potential of the proposed motion correction method for impacting downstream analysis of the estimated parametric images, improving the ability to distinguish malignant from benign hypermetabolic regions of interest. Once trained, the motion estimation inference time of our proposed network was around 460 times faster than the conventional registration baseline, showing its potential to be easily applied in clinical settings. 
\end{abstract}


\end{frontmatter}


\section{Introduction}
\label{sec1}
Whole-body positron emission tomography (PET) has been used in clinical and research protocols as a quantitative physiologic measurement primarily for oncological applications  \citep{ziegler2005positron}. Static 2-deoxy-2-[$^{18}$F]fluoro-D-glucose (FDG) imaging is a commonly applied measurement for tumor glycolytic metabolism and energy consumption \citep{muzi2012quantitative}, but semi-quantitative measurement from static PET is generally considered inferior to quantitative parameter estimation from dynamic PET, since the radiotracer uptake is a time-dependent process determined by various physiological factors \citep{karakatsanis2013dynamic}; \citep{dimitrakopoulou2021kinetic}. In dynamic PET with continuous-bed-motion mode (CBM), an image sequence with multiple frames is acquired over 90-120 minutes starting from the radiotracer injection, and voxel-by-voxel tracer kinetic modeling is applied to generate parametric images \citep{gallezot2019parametric}. The Patlak slope $K_{i}$, the net uptake rate constant, has been reported with superior tumor-to-background and contrast-to-noise ratio in oncologic lesion detection as compared to static PET imaging \citep{fahrni2019does}. 

However, the unavoidable patient motion during the long acquisition time of whole-body dynamic PET can seriously impact parametric imaging. Typically, patient motion can be further divided into respiratory motion, cardiac motion, and body motion. The intra-frame respiratory and cardiac motion introduces blurring to reconstructed image and degrades the image resolution. The inter-frame mismatch can originate from voluntary body movement and the long-term respiratory and cardiac motion pattern change, subsequently introducing attenuation correction artifacts and increasing errors in parameter estimation \citep{lu2019data}. Due to the non-rigid, complex, and unpredictable nature of patient motion, the inter-frame motion correction problem still remains challenging. Additionally, substantial tracer distribution change occurs across the dynamic frames, causing further challenges in motion estimation and correction.

Recently proposed motion correction methods for PET typically utilize external motion tracking systems, data-driven motion estimation algorithms, or traditional non-rigid registration methods. Real-time motion tracking hardware has been investigated for respiratory and head motion compensation \citep{lu2018respiratory,noonan2015repurposing}, but requires extra setup time. Data-driven joint motion estimation and correction frameworks without additional devices are preferred, which have been investigated for brain \citep{lu2020data}, respiratory \citep{feng2017self}, and body motion \citep{lu2019data}. However, these approaches have not yet been fully explored in dynamic PET, which requires taking the rapidly changing radiotracer distribution into consideration. We have previously reported that traditional non-rigid registration could successfully correct the inter-frame misalignment \citep{guo2021inter} through the entropy-based multi-resolution approach in BioImage Suite (BIS) \citep{joshi2011unified}, but the optimization process is often computationally intensive and may not be feasible in clinical translation.

Deep learning has achieved promising performance in medical image processing and analysis tasks. Convolutional neural networks (CNNs) can extract local information in images and has been widely applied in image registration. Popular baseline models include Quicksilver \citep{yang2017quicksilver} and Voxelmorph \citep{balakrishnan2019voxelmorph}, and both have been successfully implemented for brain MR image registration. However, these CNN models consider only a single image pair without distinct tracer kinetics, which could be a barrier for dynamic PET. Temporal models like recurrent neural networks (RNN) \citep{medsker2001recurrent}, including the popular long short-term memory (LSTM) networks \citep{hochreiter1997long}, have been successfully applied on sequential data analysis tasks such as language modeling \citep{merity2017regularizing}. In medical applications, LSTM models have been developed for cardio-respiratory motion prediction \citep{azizmohammadi2019model} and real-time liver tracking \citep{wang2021real}, but these works mainly focused on 1-D motion patterns. An LSTM spatial co-transformer was proposed for fetal ultrasound and magnetic resonance image registration, but was still limited to rigid motion \citep{wright2018lstm}. In combination with a CNN, the RNN and LSTM have been successfully applied in cardiac analysis such as echocardiographic assessment \citep{abdi2017quality}, automatic interpretation \citep{huang2017temporal}, and full left ventricle quantification \citep{xue2018full}.

Convolutional LSTM networks \citep{shi2015convolutional} were designed to simultaneously extract and leverage both temporal and spatial information. For video or dynamic medical image processing with 2-D or 3-D spatial features in addition to the temporal information, the combination of CNN and convolutional LSTM networks has been shown to extract spatial-temporal information effectively. In gesture and action recognition, improved accuracy and robustness were reported using a 3-D CNN with convolutional LSTM \citep{zhu2017multimodal,ge2019attention}. A spatiotemporal encoder architecture followed by a convolutional LSTM was proposed for video violence detection, gaining better performance on the  heterogeneous datasets \citep{hanson2018bidirectional}. In medical image 
applications, this combination also achieved superior performance. The cascaded convolutional LSTM layers after a CNN block achieved higher dice similarity and F-1 score in 3-D electron microscopic \citep{chen2016combining} and cardiac MR \citep{zhang2018segmentation} image segmentation. A model with a CNN and combined multiple convolutional LSTM layers was proposed for non-linear 3-D MRI-transrectal ultrasound image registration, achieving well-aligned landmarks \citep{zeng2020weakly}.

In addition to the simple concatenation following a CNN, the convolutional LSTM layer has also been used as a replacement or modification of the layers in popular CNN architectures, such as the U-Net \citep{ronneberger2015u}. The structure of using convolutional LSTM layers as the encoding layers of a U-Net has been proposed for change detection in remote sensing \citep{sun2020unet} and satellite images \citep{papadomanolaki2021deep}, and for microscopy cell segmentation \citep{arbelle2019microscopy}. The replacement of the feature map concatenations using a convolutional LSTM layer has been implemented on multimodal medical image segmentation tasks \citep{zhang2018multi,azad2019bi}. However, such combination of convolutional LSTMs with the U-net structure have yet to be fully investigated in image registration.

In dynamic PET imaging, the application of CNN-LSTM models is still nascent. To guide the selection of an optimal motion correction method for dynamic PET, a CNN-LSTM classifier was built to characterize regular and irregular breathing patterns from respiratory traces collected by an external system \citep{guo2019deep}. An automated CNN-LSTM motion correction framework was implemented for cardiac dynamic PET \citep{shi2021automatic}, but considered only rigid translation motion under simulation. A joint motion correction and denoising network was proposed with a siamese pyramid network and a bidirectional convolutional LSTM layer for low-dose gated PET \citep{zhou2021mdpet}, but focused mainly on respiratory motion. Under the whole-body scope, deep learning based inter-frame motion correction for dynamic PET still remains unexplored.

In this work, we developed a deep learning framework for unsupervised inter-frame motion correction in whole-body dynamic PET. The proposed motion estimation network is a deep CNN with an integrated convolutional LSTM layer at the bottleneck of a U-net-like architecture to capture both temporal tracer kinetics information and local spatial features simultaneously. A spatial transform layer then warps the frames according to the estimated displacement fields. We comprehensively evaluated the model performance using 9-fold cross-validation on a real patient dataset by qualitative and quantitative metrics in parametric imaging, with comparisons to both traditional and other deep learning-based registration methods. Finally, we demonstrated the potential for impacting downstream analysis of the parametric images through the classification of malignant and benign hypermetabolic regions of interest. 

\section{Material and methods}

\subsection{Dataset}
A total of 27 anonymized human subjects (5 healthy and 22 cancer patients) were included from January 2017 to February 2020 at the Yale PET center, with obtained informed consent from all the subjects and approval by the Yale Institutional Review Board. Table \ref{table1} summarizes the demographic characteristics of the included subjects. 

\begin{table}[]
\centering
\caption{Demographic characteristics of the included subjects (mean ± standard deviation, range) }
\label{table1}

\begin{tabular}{@{}ccc@{}}
\toprule
Demographic & Unit & Statistics\\ \midrule
Age & $year$ & 56 ± 14 (24 - 77)\\
Body mass index & $kg/m^2$ & 28.2 ± 3.9 (21.6 - 37.2)\\
Plasma glucose & $mg/dL$ & 101.36 ± 16.76 (75.33 - 142.00) \\
FDG injection & $mCi$ & 9.04 ± 0.88 (6.91 - 10.28)\\
\bottomrule
\end{tabular}

\end{table}

Each subject undertook a 90-min dynamic whole-body FDG CBM PET scan on a Biograph mCT PET/CT (Siemens Healthineers) with a weight-based bolus FDG injection \citep{naganawa2020assessment}. A non-contrast CT scan was acquired for attenuation correction prior to the tracer injection. In the first 6 minutes, a single-bed scan over the heart acquired 9 frames (6 × 0.5 min and 3 × 1 min) for each subject. Nineteen consecutive CBM whole-body frames were then captured (4 × 2 min and 15 × 5 min) under typewriter mode, with the bed movement always from superior to inferior and a fast shift back in each frame interval. The intra-frame bed movement speed was steady but subject-dependent as taller subjects would require faster bed motion. In 22 of 27 subjects, sequential arterial blood samples were collected and counted for plasma activity measurements. For the remaining 5 of 27 subjects, the normalized population based input function was used in parametric fitting.

All the dynamic frames were reconstructed by the ordered subsets 
expectation maximization (OSEM) algorithm \citep{hudson1994accelerated} with 21 subsets and 2 iterations on the scanner.
Attenuation, scatter, randoms, normalization, and decay corrections \citep{panin2014continuous} as well as a 5-mm 
Gaussian smoothing filter were applied to the quantitative image reconstruction following our standard clinical protocol. The voxel size of the
reconstructed images was 2.04 × 2.04 × 2.03 $mm^{3}$ for all the subjects, with the image matrix
size of 400 × 400 in the transverse plane and between 489 and 859 slices per 3D 
whole-body frame as taller patients have more slices. 57 hypermetabolic regions of interest (ROIs) (8 benign and 49 malignant) were selected in the 22 cancer subjects by a nuclear medicine physician for further motion correction evaluation.

\subsection{Network architecture}

\begin{figure*}[!t]
\centering
\includegraphics[scale=0.7]{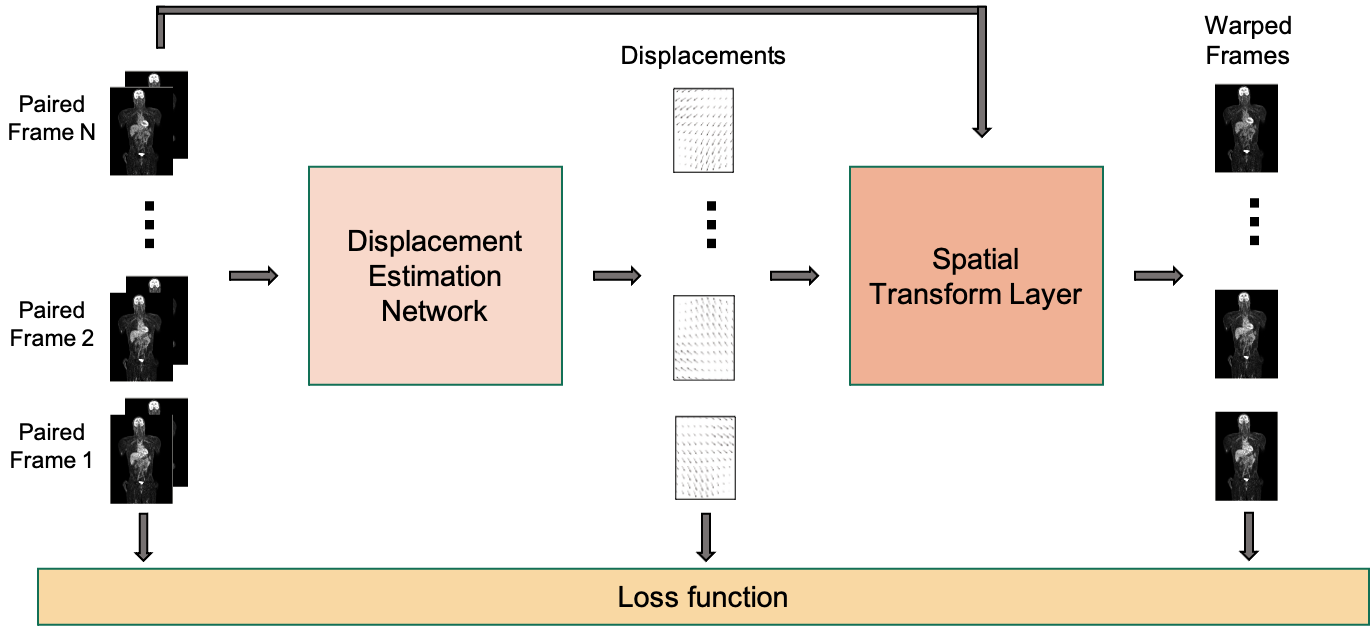}
\caption{The overall workflow of the multiple-frame motion correction framework. Each paired frame is the $i^{th}$ moving frame with the reference frame.}
\label{figure1}
\end{figure*}

In Figure \ref{figure1}, the overall workflow of the multiple-frame motion correction network is displayed. The input of the network is a dynamic frame sequence, each paired with the reference frame. With the estimated displacement fields, the following spatial transformation layer warps the original frames and get the motion compensated frames as the outputs. As in Eq. \ref{LossFunc}, the loss function of the model contains an image similarity measurement using normalized cross correlation (NCC) as well as a regularization term of the local discontinuity of the displacement fields,

\begin{equation}
\label{LossFunc}
\mathcal{L}  = \sum\limits_{j=1}^{M} -NCC\left(F_{R}, \hat{F}_{Mj}\right) + \lambda  \left|\bigtriangledown\phi_j\right|^2, 
\end{equation}
where $F_{R}$ is the reference frame, $\widehat{F_{Mj}}$ is the $j^{th}$ warped moving frame, $\lambda$ is the regularization factor, and $\phi_j$ is the estimated displacement field of the $j^{th}$ frame. The NCC similarity is superior to mean squared error when taking the rapidly changing tracer distribution across the dynamic frames into consideration, and was also shown to have better accuracy and robustness on GPU-based calculations in a previous evaluation study \citep{wu2009evaluation}.

In order to capture the tracer distribution change along time as well as the local spatial features, the motion correction network needs to be able to handle multiple dynamic frames simultaneously. LSTMs are a special kind of recurrent neural network variant that is robust and powerful for long-term dependencies \citep{shi2015convolutional}. Compared to fully connected LSTMs, the convolutional LSTMs enhanced the temporal feature extraction with spatial information encoded \citep{shi2015convolutional}. The key equations in a convolutional LSTM cell are shown in Eq. \ref{LSTM_1}-\ref{LSTM_6}, where $\circledast$ is the convolution operator and $*$ is the element-wise product:

\begin{align}
i_{t}= & {} \sigma \left( W_{i} \circledast x_{t}+U_{i}\circledast h_{t-1}+b_{i}\right) \label{LSTM_1} \\
f_{t}= & {} \sigma \left( W_{f}\circledast x_{t}+U_{f}\circledast h_{t-1}+b_{f}\right) \label{LSTM_2} \\
\tilde{c_{t}}= & {} \tanh \left( W_{c}\circledast x_{t}+U_{c}\circledast h_{t-1}+b_{c}\right) \label{LSTM_3} \\
c_{t}= & {} i_{t}*\tilde{c_{t}}+f_{t}*c_{t-1} \label{LSTM_4} \\
o_{t}= & {} \sigma \left( W_{o}\circledast x_{t}+U_{o}\circledast h_{t-1}+b_{o}\right) \label{LSTM_5} \\
h_{t}= & {} o_{t}*\tanh \left( c_{t}\right) \label{LSTM_6}
\end{align}







At the time point $t$, $x_t$ is the sequence input, $c_t$ is the cell state, $\tilde{c_{t}}$ is the estimated cell state, $c_{t-1}$ is the cell state at the previous time point $(t-1)$, $h_t$ is the hidden state, and $h_{t-1}$ is the previous hidden state. $i_t$ is the input gate (Eq. \ref{LSTM_1}) that decides whether the information from the current input will be updated to $\tilde{c_{t}}$. ${f_t}$ is the forget gate (Eq. \ref{LSTM_2}) that determines what information from $c_{t-1}$ will be kept. The current cell state is first estimated with the current input and the previous hidden state (Eq. \ref{LSTM_3}). Then, to update the cell state (Eq. \ref{LSTM_4}), the current estimation and the previous cell state will be combined with restrictions from the input and forget gates respectively. Finally, the cell state is multiplied with the output gate (Eq. \ref{LSTM_5}) to get the hidden state (Eq. \ref{LSTM_6}), which is the LSTM cell output. $W$ and $U$ matrices contain weights applied to the current input and the previous hidden state, respectively, $b$ vectors are the biases for each layer, and $\sigma$ represents the sigmoid function. In a conventional LSTM cell, the key equations are similar to a convolutional LSTM cell except all the input-to-state and state-to-state transitions are fully connected.

\begin{figure*}[!t]
\centering
\includegraphics[scale=0.63]{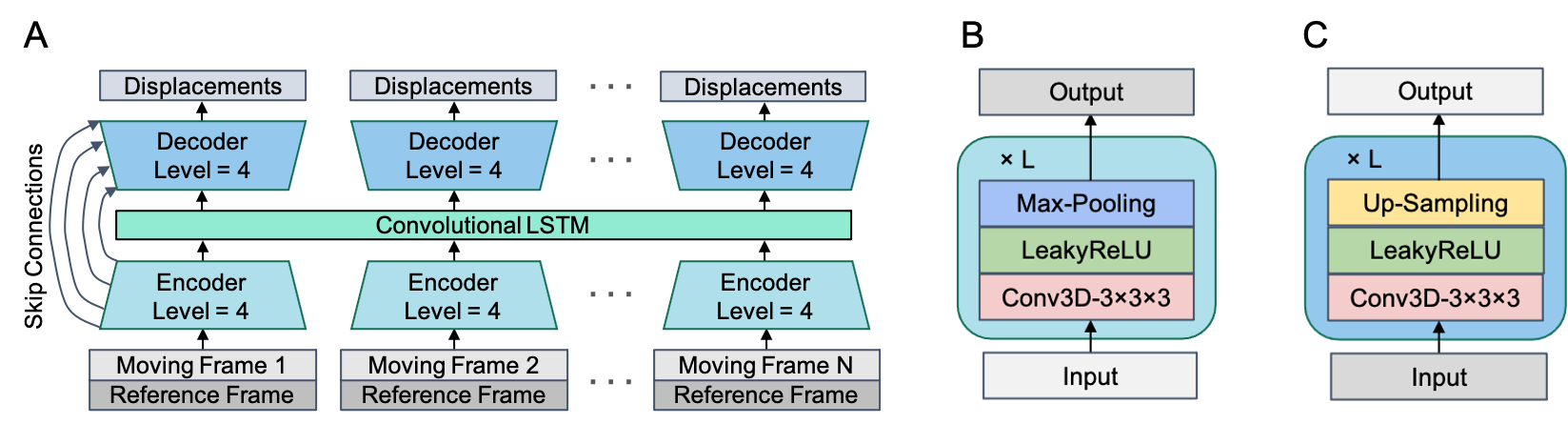}
\caption{The proposed displacement estimation network. (A) The network structure; (B) A single encoder level; (C) A single decoder level.}
\label{figure2}
\end{figure*}

Our proposed displacement estimation network is shown in Figure \ref{figure2}. We developed a convolutional motion estimation network architecture with an integrated convolutional LSTM layer, with the ability to detect and analyze multiple frames simultaneously. The spatial-temporal structure allows the network to capture the cross-frame information from not only the adjacent frames but also non-adjacent frames, especially for the long-duration motion that could affect multiple frames. In addition, capturing the inter-frame kinetic tracer distribution change can also potentially improve motion estimation accuracy. The CNN structure is similar to a 3-D U-Net \citep{cciccek20163d} with 4 encoding and decoding levels \citep{balakrishnan2019voxelmorph} and input sequence length N = 5. In each level, the 3-D convolutional layer has kernel size of 3 × 3 × 3, with 16 kernels for the first layer and 32 kernels for the following ones. The convolutional LSTM layer is combined at the bottleneck of the 3-D U-Net to process the encoded spatial-temporal features, with 32 kernels in the size of 3 × 3 × 3.

\subsection{Model training and comparison}

The inter-frame motion correction was applied to all the 5-min frames (Frame 5 - Frame 19) with Frame 12 as the reference frame. Five consecutive frames from the same subject are sent into the network at one time as the data augmentation in the temporal dimension. The input frames were downsampled by a factor of 4 and zero-padded to the same resolution of 128 × 128 × 256. All dynamic frames were converted to standardized uptake values (SUV) units prior to the network input. We implemented an intensity cutoff at SUV = 2.5 with Gaussian noise ($\sigma$ = 0.01) added to the thresholded voxels only for displacement computation but not the final images. The intensity cutoff is intended to mitigate the driving force from the high-intensity bladder voxels, and the added Gaussian noise helps avoid local saturation in the local NCC computation. The output displacement fields were then upsampled to the original resolution before warping the original frames. 

We compared our proposed model to both traditional and deep learning motion correction baselines under the same intensity cutoff implementation. The traditional entropy-based non-rigid registration is implemented in BIS \citep{joshi2011unified}, with both original and downsampled resolution images. The entropy-based normalized mutual information was chosen as the similarity measure of BIS to support the changing tracer distribution in dynamic PET with robustness in CPU-based calculations \citep{wu2009evaluation}. The optimization was conducted for two resolution levels with the final resolution at one half of the original resolution and step size as 2.0. The control point spacing of the free-form deformations was 50.0 mm with a spacing rate of 1.1 for lower resolution levels. All the other registration parameters in BioImage Suite were set as the default. The Voxelmorph framework \citep{balakrishnan2019voxelmorph} is selected as the deep learning image registration baseline. We implemented both single pairwise registration as in the original Voxelmorph framework and a variant using multiple-frame input to compare the performance difference between sequential input considering only more registration pairs and the temporal LSTM layer. In addition to the proposed convolutional LSTM combined at the U-Net bottleneck (B-convLSTM), we further tested two other variants: using a fully connected LSTM layer at the bottleneck (B-LSTM) to test the difference between dense and convolutional LSTM cells, and the serial implementation of the convolutional LSTM following the U-Net output (S-convLSTM) as comparison to the most common approach to combining CNNs with convolution LSTMs. 

All the deep learning frameworks were implemented using Keras and TensorFlow backend and trained on a NVIDIA Quadro RTX 8000 GPU. The same downsampling strategy was applied to all the deep learning based models. The stopping epoch was chosen based on the observations of reaching the minimum validation loss, which was 750 for single pairwise Voxelmorph while that of all the multiple frame models was 500. All the networks are trained using the Adam optimizer (learning rate = $10^{-4}$) with batch size = 1, the same loss function, and the regularization factor $\lambda$ = 1. 

\subsection{Parametric imaging and evaluation metrics}
The Patlak plot technique was used to analyze tracer pharmacokinetics \citep{patlak1983graphical}. As shown in Eq. \ref{PATLAK}, the Patlak plot was fitted voxel-wise with the dynamic frames and the input function after a starting time t* = 20 min \citep{ye2018improved,toczek2020accuracy},

\begin{equation}
\label{PATLAK}
\begin{aligned} 
C_T\left(t\right) = K_i \int_{0}^{t} C_P\left(\tau\right) d\tau + V_b C_P\left(t\right)
\end{aligned}
\end{equation}
where $C_T$ is the tissue tracer concentration, $C_P$ is the plasma tracer concentration (input function), the slope $K_i$ is the net uptake rate constant, and the y-axis intercept $V_b$ is the distribution volume. This approach more easily introduces image-based regression weights to the estimation process \citep{carson2005tracer}.

The $K_i$ and $V_b$ images were overlaid to qualitatively visualize any motion-related mismatch. $K_i$ and $V_b$ are sensitive to body motion \citep{lu2019data}, where $K_i$ is more heavily affected by later frames while $V_b$ is more impacted by earlier frames. Thus, better alignment of $K_i$ and $V_b$ images signifies improved registration across the dynamic frames. The whole-body normalized mutual information (NMI) as well as the whole-body NCC between $K_i$ and $V_b$ images were computed as the quantitative measurement for the alignment. 

As shown in Eq. \ref{ERROR}, the normalized weighted mean fitting errors (NFE) were computed to quantitatively assess the disparity between the dynamic data and the fitted results,
\begin{equation}
\label{ERROR}
\begin{aligned} 
NFE = \frac{\sum\limits_{k=1}^{n} w_k\left(\widehat{C_T}(t_k)-C_T(t_k)\right)^2}{\left(k-2\right)\sum\limits_{k=1}^{n}\left(\frac{w_i C_T(t_k)}{n}\right)^2}
\end{aligned}
\end{equation}
where $w_k$ is the time activity curve fitting weight after decay correction \citep{chen1991effects,chen1995decay}, $\widehat{C_T}(t_k)$ is the estimated tissue tracer
concentration, $C_T(t_k)$ is the observed tissue tracer concentration, $t_k$ represents the middle time of the $k^{th}$ frame, and n is the number of dynamic frames after t*. The voxelwise NFE maps were visualized as qualitative evaluation, and the NFE statistics in the whole-body, head, and ROIs were collected for quantitative evaluation.

A 9-fold subject-wise cross-validation was implemented to comprehensively evaluate the performance of all the motion correction methods. Significant differences between quantitative measurements from different motion correction methods were assessed using paired two-tailed t-tests  with $\alpha$ = 0.05.

\subsection{ROI analysis}
We computed the mean, maximum, and the standard deviation of the $K_i$ value for each ROI before and after each motion correction method was applied.

To assess the motion correction impact on benign/malignant classification capability, we concatenated the three reported features and trained four machine learning models for tumor classification: logistic regression (LR), linear support vector classifier (SVC), non-linear SVC with the radial basis function kernel, and random forest (RF). All the machine learning models were implemented using scikit-learn with all the default hyperparameters. 
As the ROI group is relatively small and imbalanced with more malignant lesions than benign ones, a 5-fold stratified cross validation was implemented with assigned sample weights inversely proportional to the benign/malignant ratio. For each machine learning model, the mean receiver operating characteristic (ROC) curve was generated from the mean of the sensitivity and specificity values for the 5 folds by varying the classification score thresholds, and the area under the curve (AUC) was then computed as the assessment of classification capability.

\section{Results}
\subsection{Qualitative evaluations}

\begin{figure*}[!t]
\centering
\includegraphics[scale=0.53]{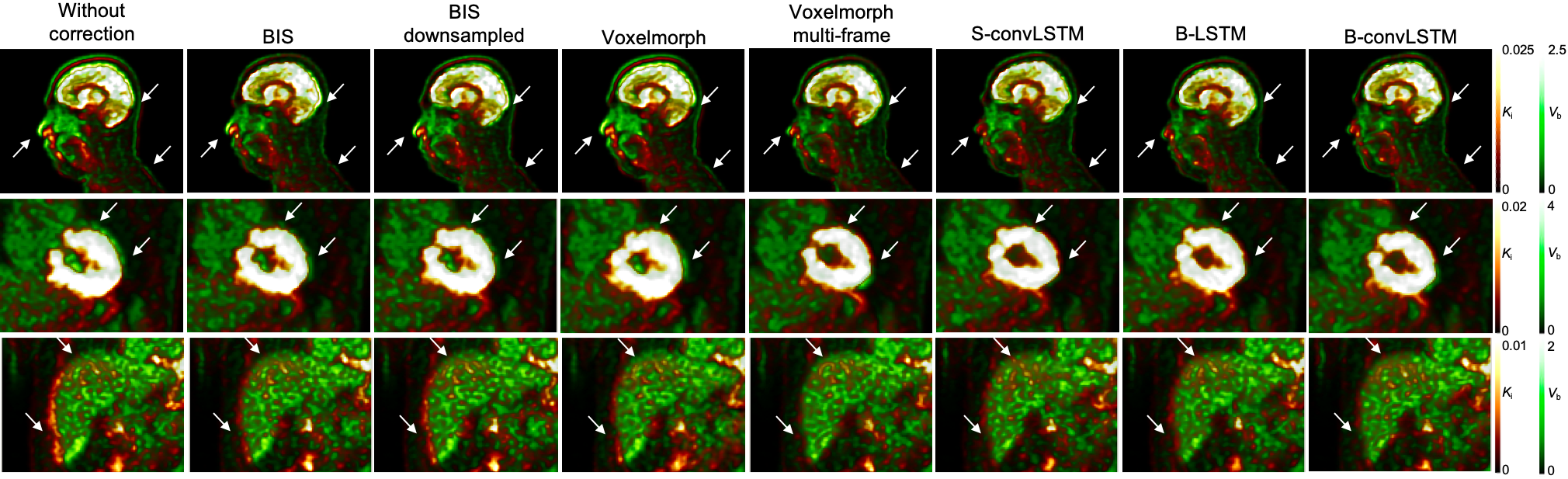}
\caption{Sample overlaid Patlak $K_i$(red) and $V_b$(green) images showing inter-frame motion and correction impacts in brain (upper), heart (middle), and liver (bottom).}
\label{figure3}
\end{figure*}

Figure \ref{figure3} shows the inter-frame motion impact and the correction effects on the overlaid Patlak $K_i$ and $V_b$ images. Without motion correction, the Patlak $K_i$ and $V_b$ overlaid images display spatial misalignments originating from the inter-frame motion. The non-rigid method of BIS substantially reduced the misalignment at the head, heart, and liver, but under-correction existed with some remaining mismatch. The performance of BIS further degraded when operating on downsampled dynamic frames, which were the same inputs used for the deep learning approaches. The baseline model Voxelmorph with single image pair input also showed similar trends, with improved alignment but remaining mismatch as well. The multi-frame version of Voxelmorph further enhanced the $K_i$ and $V_b$ alignment, showing that letting the network "see" more input frames simultaneously is a key improvement in the dynamic frame registration. The results of the B-LSTM model were even better aligned, indicating the power of recurrent analysis. Finally, the two convolutional LSTM models achieved the best visual spatial alignment improvement. Specifically, the B-convLSTM model corrected the non-rigid misalignment at the lower right brain edge and the liver boundary the best, while the S-convLSTM model reduced the misalignment from the rigid motion at the nose outline and the heart boundary the most. Note that the heart underwent significant motion and all the motion correction methods changed the spatial location and the shape of the heart, with the B-convLSTM model preserving the heart shape the best. This implies that motion correction using the multi-frame input and the convolutional LSTM for extracting spatial and temporal information could enhance the qualitative spatial alignment of the parametric images. 

The significant motion impact on $K_i$ and $V_b$ misalignment could be detected in 23 out of 27 subjects; the most commonly detected voluntary movement of the head including rigid movement of the skull and the brain deformation could be found in all of the 27 subjects. Some motion-related artifacts include the "ridge-valley" artifact \citep{lu2019data} and the same hypermetabolic peak appearing in adjacent locations in $K_i$ and $V_b$ images. According to the sample figures, the proposed spatial-temporal network was able to further reduce those artifacts and improve the motion compensation as well as lead to altered $K_i$ and $V_b$ values, though the ground truth of the $K_i$ and $V_b$ values are lacking in a real patient dataset.

\begin{figure*}[!h]
\centering
\includegraphics[scale=0.54]{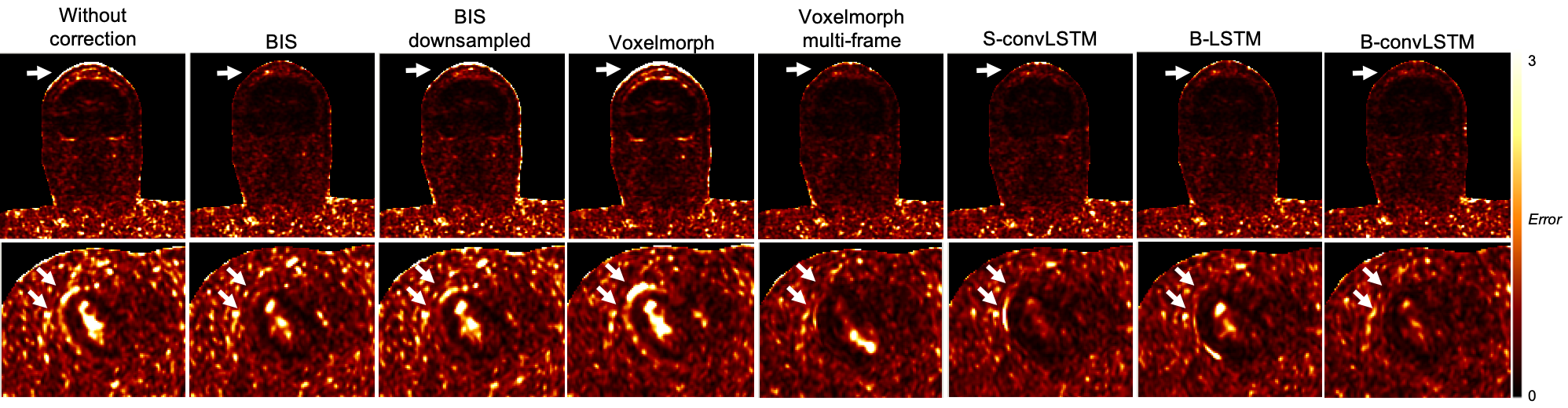}
\caption{Sample voxel-wise Patlak NFE maps of skull and brain (upper) and heart (bottom).}
\label{figure4}
\end{figure*}

The sample voxel-wise NFE maps are visualized in Figure \ref{figure4}. The NFE map of the proposed network was overall darker than the maps without motion correction and other motion correction baselines under the same color bar scaling, indicating further reduced fitting error in the Patlak model. The total number of hotspot voxels representing higher voxel-wise fitting errors decreased in the sample NFE map after the proposed motion correction. For the most common head motion, the proposed networks further reduced the high error peaks at the head and brain edge. For the heart region with high fitting error due to the significant motion, the proposed networks additionally diminished the hotspots in the ventricles and at the edges. Our proposed model B-convLSTM reduced the peaks and the brightness of the NFE maps the most.

\subsection{Quantitative evaluations}

\begin{table*}[]
\small
\newcommand{\tabincell}[2]{\begin{tabular}{@{}#1@{}}#2\end{tabular}}
\centering
\caption{Quantitative assessments of the inter-frame motion correction methods (mean ± standard deviation, range) }
\label{table2}

\begin{tabular}{@{}lcccccc@{}}
\toprule
 &\tabincell{c}{Whole-body\\mean NFE} & \tabincell{c}{Whole-body\\maximum NFE} & Head NFE & ROI NFE & $K_i$/$V_b$ NMI & $K_i$/$V_b$ NCC  \\  \midrule
\specialrule{0em}{2pt}{2pt}
\tabincell{l}{Without \\correction}     & \tabincell{c}{0.1559 ± 0.0696 \\ (0.0716 - 0.4093)} & \tabincell{c}{16.1518 ± 3.6710 \\(7.2930 - 22.2021)}     & \tabincell{c}{0.1103 ± 0.0530 \\(0.0495 - 0.2630)} & \tabincell{c}{1.3526 ± 2.1897 \\(0.1831 - 15.9933)} & \tabincell{c}{0.9298 ± 0.0312 \\(0.8567 - 0.9738) }& \tabincell{c}{0.2695 ± 0.1110 \\(0.0503 - 0.4512)} \\
\specialrule{0em}{2pt}{2pt}
BIS & \tabincell{c}{0.1418 ± 0.0584\\ (0.0686 - 0.3671)} & \tabincell{c}{70.1237 ± 245.5716 \\(7.2911 - 1314.6486)} & \tabincell{c}{0.0915 ± 0.0442 \\(0.0438 - 0.2479)} & \tabincell{c}{1.0298 ± 1.1165 \\(0.1571 - 7.4314)}  & \tabincell{c}{0.9451 ± 0.0283 \\(0.8567 - 0.9794)} & \tabincell{c}{0.3150 ± 0.1067 \\(0.1179 - 0.5354)} \\
\specialrule{0em}{2pt}{2pt}
\tabincell{l}{BIS\\downsampled} & \tabincell{c}{0.1517 ± 0.0597 \\(0.0730 - 0.3725)}& \tabincell{c}{87.8017 ± 245.8142 \\(4.7684 - 1003.5248)} &\tabincell{c}{ 0.0980 ± 0.0446 \\(0.0513 - 0.2594) }& \tabincell{c}{1.1361 ± 1.6262 \\(0.1647 - 11.8469)} & \tabincell{c}{0.9430 ± 0.0248 \\(0.8783 - 0.9736) }& \tabincell{c}{0.3151 ± 0.1007 \\(0.1336 - 0.5079) }\\
\specialrule{0em}{2pt}{2pt}
Voxelmorph & \tabincell{c}{0.1492 ± 0.0640 \\(0.0670 - 0.3865)} & \tabincell{c}{15.2463 ± 4.0112 \\(7.2963 - 21.9768)  }   & \tabincell{c}{0.1152 ± 0.0463\\ (0.0517 - 0.2480)} &\tabincell{c}{ 1.3616 ± 2.0005 \\(0.2769 - 14.4634) }&\tabincell{c}{ 0.9319 ± 0.0303 \\(0.8601 - 0.9748)} & \tabincell{c}{0.2733 ± 0.1109 \\(0.0525 - 0.4496)} \\
\specialrule{0em}{2pt}{2pt}
\tabincell{l}{Voxelmorph \\multi-frame} & \tabincell{c}{0.1181 ± 0.0476\\ (0.0548 - 0.3023)} &\tabincell{c}{ 15.2255 ± 4.0066 \\(3.9533 - 21.5372) }    & \tabincell{c}{0.0801 ± 0.0385 \\(0.0372 - 0.2128)} & \tabincell{c}{0.8949 ± 1.1831\\ (0.1494 - 8.4020) } & \tabincell{c}{0.9540 ± 0.0219 \\(0.8990 - 0.9828) }& \tabincell{c}{0.5021 ± 0.1639\\ (0.1557 - 0.7245)} \\
\specialrule{0em}{2pt}{2pt}
S-convLSTM  & \tabincell{c}{0.1180 ± 0.0474 \\(0.0573 - 0.2982)} & \tabincell{c}{14.9111 ± 4.1132 \\(4.9149 - 21.5393) }    & \tabincell{c}{0.0798 ± 0.0384\\ (0.0378 - 0.2131)} & \tabincell{c}{0.9289 ± 1.4284\\ (0.1472 - 10.5453)} & \tabincell{c}{\textbf{0.9547 ± 0.0212}\\ \textbf{(0.9010 - 0.9832)}} &\tabincell{c}{ \textbf{0.5141 ± 0.1524} \\\textbf{(0.2156 - 0.7644)} }\\
\specialrule{0em}{2pt}{2pt}
B-LSTM & \tabincell{c}{0.1186 ± 0.0488\\ (0.0552 - 0.3039) }& \tabincell{c}{15.2877 ± 3.6730 \\(6.5119 - 21.3882)}     & \tabincell{c}{0.0797 ± 0.0389 \\(0.0375 - 0.2159) }& \tabincell{c}{1.0814 ± 2.3849 \\(0.1353 - 18.0843)} & \tabincell{c}{0.9539 ± 0.0221 \\(0.8997 - 0.9829) }& \tabincell{c}{0.4963 ± 0.1587\\ (0.1295 - 0.7371)} \\
\specialrule{0em}{2pt}{2pt}
B-convLSTM & \tabincell{c}{\textbf{0.1171 ± 0.0476} \\\textbf{(0.0547 - 0.2994)}} & \tabincell{c}{\textbf{14.4556 ± 4.3182} \\\textbf{(4.4162 - 21.2903)}}    & \tabincell{c}{\textbf{0.0792 ± 0.0383} \\\textbf{(0.0375 - 0.2122)}}& \tabincell{c}{\textbf{0.8690 ± 1.0808} \\\textbf{(0.1521 - 7.6837)} } &\tabincell{c}{ 0.9541 ± 0.0223\\ (0.8970 - 0.9832)} & \tabincell{c}{0.5059 ± 0.1539 \\(0.1538 - 0.7691)}\\
\bottomrule
\end{tabular}
\end{table*}

Table \ref{table2} summarizes the quantitative evaluations of the inter-frame motion correction methods. The proposed B-convLSTM network significantly reduced mean NFE in Patlak fitting in the whole-body and subareas with significant motion (head and ROIs) and raised the quantitative $K_i$/$V_b$ alignment measurements compared with traditional non-rigid and Voxelmorph baselines (all with $p < 0.05$). For the whole-body maximum NFE, the BIS methods tended to increase the peak values, with the average maximum NFE increasing by $\sim$5 fold and the largest errors increasing by $\sim$50 fold. In contrast, the proposed model successfully reduced both the lower and upper end of the peak error range. The B-convLSTM model achieved the lowest whole-body, head, and ROI NFE while the S-convLSTM attained the highest $K_i$/$V_b$ NMI and NCC, potentially indicating that the S-convLSTM model is more sensitive to matching image noise.

The inference (wall) time of motion estimation and the model size of each motion correction method are summarized in Table \ref{comparison}. Note that all the deep learning models are trained under the downsampled resolution while the original BIS method used the full resolution images. Under the same downsampled resolution, the well-trained B-convLSTM model could perform motion estimation $\sim$17 times faster than BIS and $\sim$6 times faster than the single image pair Voxelmorph, since the multiple frame model is able to estimate the displacements in parallel. Although the multi-frame Voxelmorph model had the lowest inference time and number of parameters, the motion correction performance was significantly worse than that of the proposed B-convLSTM (Table \ref{table2}). In contrast, while there were no further significant differences ($p > 0.05$) between the motion correction results of S-convLSTM and B-convLSTM, the time consumption of B-convLSTM was 11.9\% lower than that of S-convLSTM with similar model size. Finally, the time consumption of B-convLSTM is only 61\% of the B-LSTM model with considerably reduced model size, showing that in additional to better utilizing spatial-temporal information, another advantage of the bottleneck convolutional LSTM layer is significantly reducing the time and memory consumption.

Since the BIS motion correction for each pair of frames were run in parallel, we also measured the overall compute time for a fair comparison. The total compute time includes the spatial transformation time and the upsampling time for the models under the downsampled resolution in addition to the motion estimation inference time. With the displacement field upsampling and spatial warping, the overall compute time of B-convLSTM is $\sim$65.5 min per subject (4.65 min per frame), which is still $\sim$12 times faster than the fully parallelized original resolution BIS running 14 jobs in parallel ($\sim$13.3 hr). Without the parallel compute resource, the original BIS would take $\sim$1 week and the downsampled BIS would need $\sim$7.3 hours to process one subject, showing the great advantage of the savings in compute time for B-convLSTM.

\begin{table}[]
\small
\newcommand{\tabincell}[2]{\begin{tabular}{@{}#1@{}}#2\end{tabular}}
\centering
\caption{The inference time of motion estimation and the model size comparison. }
\label{comparison}

\begin{tabular}{@{}lcc@{}}
\toprule
 &\tabincell{c}{Motion estimation inference \\time of one subject (second)} & \tabincell{c}{Trainable parameters}   \\  \midrule
\specialrule{0em}{2pt}{2pt}

BIS & 48034 & --  \\
\specialrule{0em}{2pt}{2pt}
\tabincell{l}{BIS\\downsampled} & 1813 & -- \\
\specialrule{0em}{2pt}{2pt}
Voxelmorph & 642 & 327,331  \\
\specialrule{0em}{2pt}{2pt}
\tabincell{l}{Voxelmorph \\multi-frame} & 101 & 327,331     \\
\specialrule{0em}{2pt}{2pt}
S-convLSTM  & 118 & 501,571   \\
\specialrule{0em}{2pt}{2pt}
B-LSTM & 170 & 138,744,355    \\
\specialrule{0em}{2pt}{2pt}
B-convLSTM & 104 & 548,643   \\
\bottomrule
\end{tabular}
\end{table}

\subsection{ROI analysis and malignancy classification}

The $K_i$ statistics of the benign and malignant ROIs was listed in Table \ref{table3}. All the motion correction methods suggested the general trend of reducing the mean and maximum $K_i$ values while increasing the standard deviation of $K_i$ in each ROI, which matches with the motion impact found in the simulation results as shown in Supplementary Table \ref{SupTable1}. The deep learning motion correction results further enhanced this impact compared with BIS, and the two convolutional LSTM networks showed similar amount of substantial change.

\begin{table*}[]
\small
\newcommand{\tabincell}[2]{\begin{tabular}{@{}#1@{}}#2\end{tabular}}
\centering
\caption{$K_i$ statistics of the benign and malignant ROI groups for each inter-frame motion correction method (mean ± standard deviation, range) }
\label{table3}

\begin{tabular}{@{}lcccccc@{}}
\toprule
 &  \multicolumn{3}{c}{Benign (n = 8)} & \multicolumn{3}{c}{Malignant (n = 49)}\\
\specialrule{0em}{1pt}{1pt}
\cline{2-7}
\specialrule{0em}{2pt}{2pt}
 & Mean & Maximum & Standard deviation & Mean & Maximum & Standard deviation  \\  \midrule
\specialrule{0em}{2pt}{2pt}
\tabincell{l}{Without \\correction}     & \tabincell{c}{0.0050 ± 0.0028 \\ (0.0017 - 0.0107)} & \tabincell{c}{0.0083 ± 0.0044 \\(0.0024 - 0.0173)}     & \tabincell{c}{0.0015 ± 0.0007 \\(0.0003 - 0.0027)} & \tabincell{c}{0.0138 ± 0.0079 \\(0.0009 - 0.0392)} & \tabincell{c}{0.0288 ± 0.0192 \\(0.0017 - 0.0835) }& \tabincell{c}{0.0051 ± 0.0036 \\(0.0002 - 0.0142)} \\
\specialrule{0em}{2pt}{2pt}
BIS & \tabincell{c}{0.0049 ± 0.0028\\ (0.0016 - 0.0108)} & \tabincell{c}{0.0083 ± 0.0046 \\(0.0022 - 0.0181)} & \tabincell{c}{0.0014 ± 0.0007 \\(0.0003 - 0.0025)} & \tabincell{c}{0.0134 ± 0.0077 \\(0.0010 - 0.0383)}  & \tabincell{c}{0.0287 ± 0.0187 \\(0.0018 - 0.0770)} & \tabincell{c}{0.0053 ± 0.0037 \\(0.0002 - 0.0141)} \\
\specialrule{0em}{2pt}{2pt}
\tabincell{l}{BIS\\downsampled} & \tabincell{c}{0.0049 ± 0.0028 \\(0.0016 - 0.0110)}& \tabincell{c}{0.0082 ± 0.0045 \\(0.0023 - 0.0179)} &\tabincell{c}{ 0.0015 ± 0.0007 \\(0.0003 - 0.0026) }& \tabincell{c}{0.0132 ± 0.0078 \\(0.0008 - 0.0368)} & \tabincell{c}{0.0283 ± 0.0186 \\(0.0019 - 0.0758) }& \tabincell{c}{0.0052 ± 0.0037 \\(0.0002 - 0.0140) }\\
\specialrule{0em}{2pt}{2pt}
Voxelmorph & \tabincell{c}{0.0043 ± 0.0031 \\(0.0011 - 0.0116)} & \tabincell{c}{0.0076 ± 0.0045 \\(0.0019 - 0.0173)  }   & \tabincell{c}{0.0015 ± 0.0007\\ (0.0003 - 0.0025)} &\tabincell{c}{ 0.0119 ± 0.0074 \\(0.0010 - 0.0382) }&\tabincell{c}{0.0271 ± 0.0183 \\(0.0015 - 0.0827)} & \tabincell{c}{0.0053 ± 0.0039 \\(0.0003 - 0.0159)} \\
\specialrule{0em}{2pt}{2pt}
\tabincell{l}{Voxelmorph \\multi-frame} & \tabincell{c}{0.0047 ± 0.0024\\ (0.0013 - 0.0094)} &\tabincell{c}{0.0082 ± 0.0045 \\(0.0018 - 0.0168) }    & \tabincell{c}{0.0017 ± 0.0009 \\(0.0003 - 0.0031)} & \tabincell{c}{0.0116 ± 0.0071\\ (0.0003 - 0.0344) } & \tabincell{c}{0.0276 ± 0.0186 \\(0.0012 - 0.0743) }& \tabincell{c}{0.0055 ± 0.0039\\ (0.0001 - 0.0150)} \\
\specialrule{0em}{2pt}{2pt}
S-convLSTM  & \tabincell{c}{0.0041 ± 0.0027 \\(0.0008 - 0.0101)} & \tabincell{c}{0.0076 ± 0.0044 \\(0.0015 - 0.0171) }    & \tabincell{c}{0.0014 ± 0.0007\\ (0.0003 - 0.0028)} & \tabincell{c}{0.0118 ± 0.0070\\ (0.0008 - 0.0338)} & \tabincell{c}{0.0276 ± 0.0186\\ (0.0013 - 0.0734)} &\tabincell{c}{0.0055 ± 0.0039 \\(0.0002 - 0.0154) }\\
\specialrule{0em}{2pt}{2pt}
B-LSTM & \tabincell{c}{0.0039 ± 0.0021\\ (0.0014 - 0.0087) }& \tabincell{c}{0.0075 ± 0.0045 \\(0.0021 - 0.0172)}     & \tabincell{c}{0.0017 ± 0.0010 \\(0.0003 - 0.0033) }& \tabincell{c}{0.0115 ± 0.0067 \\(0.0007 - 0.0333)} & \tabincell{c}{0.0275 ± 0.0187 \\(0.0012 - 0.0725) }& \tabincell{c}{0.0056 ± 0.0041\\ (0.0002 - 0.0158)} \\
\specialrule{0em}{2pt}{2pt}
B-convLSTM & \tabincell{c}{0.0041 ± 0.0024 \\(0.0009 - 0.0091)} & \tabincell{c}{0.0075 ± 0.0045 \\(0.0013 - 0.0172)}    & \tabincell{c}{0.0015 ± 0.0010 \\(0.0002 - 0.0033)}& \tabincell{c}{0.0118 ± 0.0068 \\(0.0003 - 0.0335)} &\tabincell{c}{0.0277 ± 0.0183\\ (0.0009 - 0.0742)} & \tabincell{c}{0.0054 ± 0.0039 \\(0.0002 - 0.0156)}\\
\bottomrule
\end{tabular}
\end{table*}

The AUCs of all the machine learning classifiers trained for hypermetabolic ROI malignancy classification from local $K_i$ statistics was summarized in Table \ref{table4}. The average of the four (two linear and two non-linear) machine learning algorithms was also computed to assess the overall ability of distinguishing benign/malignant ROIs based on the $K_i$ statistics. The B-convLSTM model reached the highest average AUC with lowest standard deviation and performed best for two of the four learning methods, indicating the improved classification capability compared to other motion correction methods. Note that the ROI group is relatively small and imbalanced since all the hotspots were chosen in the cancer subjects by the physician and some of those were later categorized as benign. The average ROC curve of each inter-frame motion correction method was plotted in Supplementary Figure \ref{SupFig1}, where the curve of the B-convLSTM model exceeded that of all the other motion correction methods.

\begin{table*}[]
\newcommand{\tabincell}[2]{\begin{tabular}{@{}#1@{}}#2\end{tabular}}
\centering
\caption{The AUCs of all the machine learning ROI malignancy classifiers for each motion correction method (mean ± standard deviation) }
\label{table4}

\begin{tabular}{@{}lccccc@{}}
\toprule

\specialrule{0em}{1pt}{1pt}
 & SVC & SVC Linear & RF & LR & Average  \\  \midrule
\specialrule{0em}{2pt}{2pt}
\tabincell{l}{Without \\correction}     & 0.8500 ± 0.1483 & 0.8500 ± 0.1483    & 0.6856 ± 0.2855 &0.8500 ± 0.1483 & 0.8008 ± 0.2151 \\
\specialrule{0em}{2pt}{2pt}
BIS & \textbf{0.8900 ± 0.1114} & 0.8700 ± 0.1166 & 0.6906 ± 0.2669 & 0.8700 ± 0.1166 & 0.8264 ± 0.2018 \\
\specialrule{0em}{2pt}{2pt}
\tabincell{l}{BIS\\downsampled} & 0.8600 ± 0.1200 & 0.8900 ± 0.1114
 & 0.7578 ± 0.2486 & 0.8900 ± 0.1114 & 0.8519 ± 0.1599 \\
\specialrule{0em}{2pt}{2pt}
Voxelmorph & 0.8100 ± 0.1855 & 0.8300 ± 0.1833 & 0.7317 ± 0.2172 & 0.8300 ± 0.1833 & 0.8004 ± 0.1971 \\
\specialrule{0em}{2pt}{2pt}
\tabincell{l}{Voxelmorph \\multi-frame} & 0.8700 ± 0.1470 & 0.8500 ± 0.1483   & 0.8428 ± 0.1427 & 0.8500 ± 0.1483
 & 0.8532 ± 0.1470 \\
\specialrule{0em}{2pt}{2pt}
S-convLSTM  & 0.8478 ± 0.1086 & 0.8878 ± 0.0661  & 0.8528 ± 0.1045 & 0.8878 ± 0.0661 & 0.8640 ± 0.0982\\
\specialrule{0em}{2pt}{2pt}
B-LSTM & 0.8589 ± 0.1059 & 0.8589 ± 0.1059  & \textbf{0.9228 ± 0.0404} & 0.8589 ± 0.1059 & 0.8719 ± 0.0998 \\
\specialrule{0em}{2pt}{2pt}
B-convLSTM & 0.8789 ± 0.1022 & \textbf{0.9100 ± 0.0800} & 0.9100 ± 0.1114 & \textbf{0.9100 ± 0.0800} & \textbf{0.9022 ± 0.0954}\\
\bottomrule
\end{tabular}
\end{table*}

\begin{figure*}[]
\centering
\includegraphics[scale=0.54]{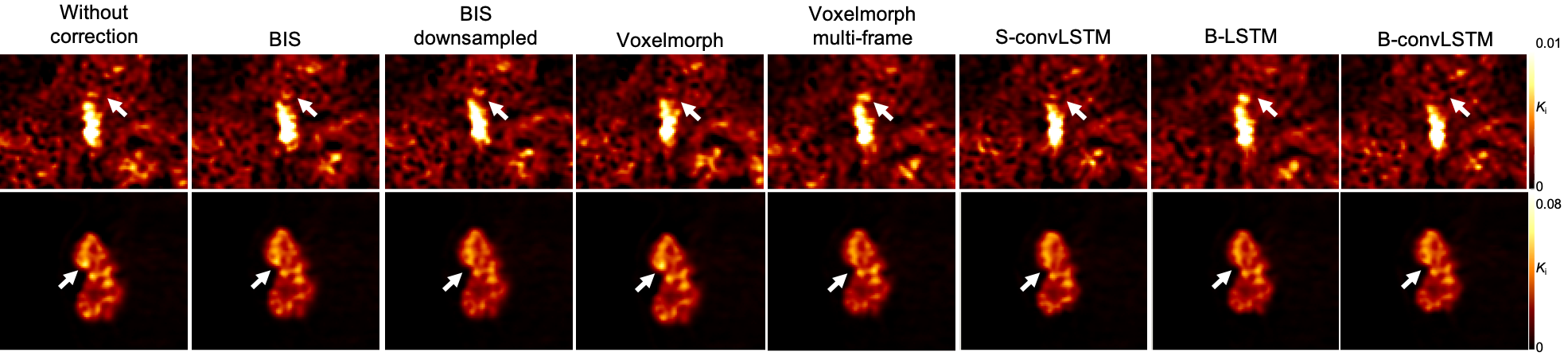}
\caption{Sample hypermetabolic ROI $K_i$ images after each inter-frame motion correction method.}
\label{figure5}
\end{figure*}

The sample hypermetabolic ROI $K_i$ images are displayed in Figure \ref{figure5}. The proposed inter-frame motion correction provided additional location, edge and texture changes compared to other baselines. For the lesion of esophageal adenocarcinoma (Figure \ref{figure5} upper), the small artifact-like hotspot above the major lesion had a lower peak after the BIS correction, and the proposed B-convLSTM model further reduced its shape and values, in addition to sharpening and enhancing the edge of the lesion. The effect on motion-related artifacts also matches with the tumor simulation results in Supplementary Figure \ref{SupFig1}. For the metastatic melanoma lesion (Figure \ref{figure5} lower), the traditional baseline BIS reduced the peaks of the hot spots and amplified the tumor texture. However, the Voxelmorph deep learning baseline increased the brightness of the originally strongest hot spot. After the motion correction of the B-convLSTM model, the estimations of peak $K_i$ values further decreased compared to BIS. The visual differences after the proposed motion correction network might contribute to the improvement of clinical diagnosis and malignancy discrimination, although it is noted that gold standard motion-free $K_i$ images are not available for the real patient dataset.

\subsection{Hyperparameter sensitivity test}

A sensitivity test was run to test the impact of regularization term lambda on the displacement field local discontinuity. According to Table \ref{table5}, $\lambda$ = 1 gives the lowest NFE while $\lambda$ = 0.1 reaches the highest $K_i$/$V_b$ NMI and NCC. As shown in Figure \ref{figure6}, $\lambda$ = 1 performed the best on $K_i$ and $V_b$ spatial overlay with the least mismatch. It's likely that the weakest penalization $\lambda$ = 0.1 also matched the image noise, which caused higher intensity-based alignment but also higher NFE and more visual mismatch. With a stronger regularization ($\lambda > 1$), the model is also less effective in motion estimation. The optimal $\lambda$ was set to 1, consistent with the suggestions from the original Voxelmorph study \citep{balakrishnan2019voxelmorph}.

\begin{table*}[!ht]
\small
\newcommand{\tabincell}[2]{\begin{tabular}{@{}#1@{}}#2\end{tabular}}
\centering
\caption{Quantitative assessments of the sensitivity test (mean ± standard deviation, range) }
\label{table5}

\begin{tabular}{@{}lcccccc@{}}
\toprule
 &\tabincell{c}{Whole-body\\mean NFE} & \tabincell{c}{Whole-body\\maximum NFE} & Head NFE & ROI NFE & $K_i$/$V_b$ NMI & $K_i$/$V_b$ NCC  \\  \midrule
\specialrule{0em}{2pt}{2pt}
$\lambda$ = 0.1 & \tabincell{c}{0.1183 ± 0.0480 \\ (0.0567 - 0.3067)} & \tabincell{c}{15.4661 ± 4.0306 \\(4.5706 - 21.3611)}     & \tabincell{c}{0.0832 ± 0.0387 \\(0.0394 - 0.2187)} & \tabincell{c}{1.0495 ± 2.0207 \\(0.1789 - 15.2915)} & \tabincell{c}{\textbf{0.9561 ± 0.0211} \\\textbf{(0.9017 - 0.9839)}}& \tabincell{c}{\textbf{0.5198 ± 0.1626} \\\textbf{(0.1866 - 0.8033)}} \\
\specialrule{0em}{2pt}{2pt}
$\lambda$ = 1 & \tabincell{c}{\textbf{0.1171 ± 0.0476}\\ \textbf{(0.0547 - 0.2994)}} & \tabincell{c}{\textbf{14.4556 ± 4.3182} \\\textbf{(4.4162 - 21.2903)}} & \tabincell{c}{\textbf{0.0792 ± 0.0383} \\\textbf{(0.0375 - 0.2122)}} & \tabincell{c}{\textbf{0.8690 ± 1.0808} \\\textbf{(0.1521 - 7.6837)}}  & \tabincell{c}{0.9541 ± 0.0223 \\(0.8970 - 0.9832)} & \tabincell{c}{0.5059 ± 0.1539 \\(0.1538 - 0.7691)} \\
\specialrule{0em}{2pt}{2pt}
$\lambda$ = 10 & \tabincell{c}{0.1239 ± 0.0523 \\(0.0576 - 0.3188)}& \tabincell{c}{14.5822 ± 4.4000 \\(4.2601 - 21.6528)} &\tabincell{c}{ 0.0821 ± 0.0391 \\(0.0380 - 0.2160) }& \tabincell{c}{1.0364 ± 1.8285 \\(0.1373 - 13.8209)} & \tabincell{c}{0.9492 ± 0.0237 \\(0.8930 - 0.9820) }& \tabincell{c}{0.4384 ± 0.1408 \\(0.1547 - 0.7410) }\\
\specialrule{0em}{2pt}{2pt}
$\lambda$ = 100 & \tabincell{c}{0.1342 ± 0.0576 \\(0.0644 - 0.3466)} & \tabincell{c}{15.0293 ± 4.4169 \\(6.9871 - 21.9813)  }   & \tabincell{c}{0.0906 ± 0.0434\\ (0.0410 - 0.2311)} &\tabincell{c}{1.0449 ± 1.2975 \\(0.1917 - 9.1753) }&\tabincell{c}{0.9395 ± 0.0275 \\(0.8724 - 0.9784)} & \tabincell{c}{0.3182 ± 0.1101 \\(0.1094 - 0.5402)} \\

\bottomrule
\end{tabular}
\end{table*}

\begin{figure*}[!h]
\centering
\includegraphics[scale=0.6]{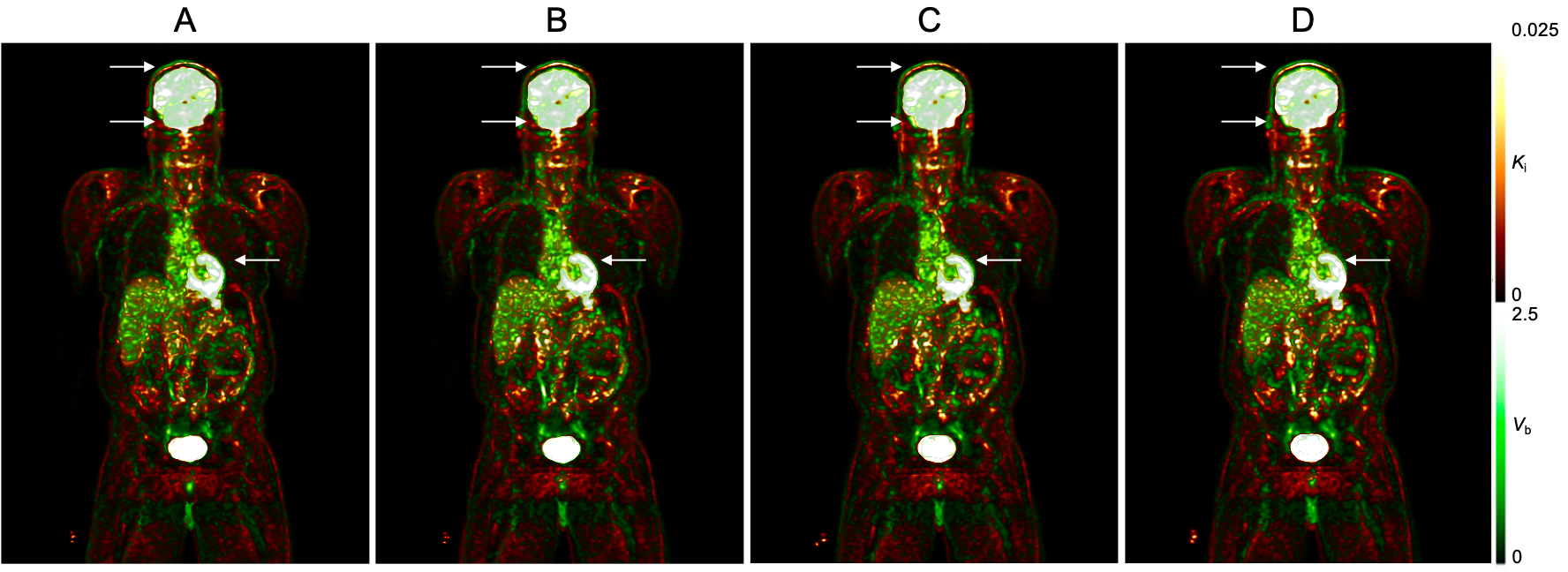}
\caption{The sample overlaid $K_i$ (red) / $V_b$ (green) images of the sensitivity test. (A) $\lambda$ = 0.1; (B) $\lambda$ = 1; (C) $\lambda$ = 10; (D) $\lambda$ = 100.}
\label{figure6}
\end{figure*}


\section{Discussion}

In this work, we proposed the usage of convolutional LSTM integrated into a deep convolutional network for fast inter-frame motion correction in dynamic PET and parametric imaging. The network allows the model to "see" multiple dynamic image series as the input together and process the spatial-temporal information and features. Compared to traditional and deep learning baselines, the proposed method successfully further enhanced the spatial alignment of the parametric $K_i$ and $V_b$ images and reduced inter-frame mismatch. The Patlak NFE was significantly reduced in the whole-body and head and ROI subareas, which exhibited considerable motion. The further modifications of the ROI appearances were observed with greater texture and shape change, and the potential of further improving the benign/malignant classification capacity following motion correction with our B-convLSTM model was suggested. Once trained, it takes $\sim$4.65 min/frame to apply the proposed motion correction model, substantially reducing the time consumption compared with the traditional BIS method ($\sim$13.3 hr/frame). 

A major obstacle holding back the clinical application of BIS motion correction is the needed long computing time. Using downsampled dynamic frames as the input was able to speed up the process ($\sim$31 min/frame), but the registration performance also degraded due to the lower resolution. Due to the memory limitation, the proposed motion correction network was inputted with the downsampled dynamic frame series, and the results computed based on a lower resolution still achieved greater qualitative and quantitative performances than the full-resolution BIS. The length of the input frame sequences for the proposed model was fixed at 5, which was limited by the GPU memory resource, although the high memory and computational resource requirement are from the 4-D whole-body data. Therefore, the network has the potential to be extended for longer time sequences with lower resolution such as cardiac motion tracking/prediction.

Compared to the deep learning baseline Voxelmorph with one single image pair, the Voxelmorph multi-frame showed that it is promising to have sequential inputs and the multiple frame pair registration for the dynamic PET images with related inter-frame kinetic information. We then demonstrated the advantage of integrating an LSTM layer into the multiple frame network by experimenting with three models: S-convLSTM, with a convolutional LSTM layer serially following the multi-frame Voxelmorph network; B-LSTM, with a standard fully-connected LSTM layer implemented at the bottleneck of the U-net in the multiframe Voxelmorph network; and our proposed B-convLSTM, with a convolutional LSTM layer integrated into the bottleneck of the U-net. The input feature map size for models with sequential connections is considerably larger than the bottleneck integration, making an S-LSTM (serial fully connected LSTM following the U-Net output) intractable, and thus only B-LSTM was tested. Compared with the fully connected LSTM, the convolutional LSTM showed its superiority in extracting spatial and temporal information simultaneously from the 4-D image series while also saving the computation and memory resources. The two ways of integrating the convolutional LSTM layer, i.e., S-convLSTM and B-convLSTM, both showed advantages in some aspects. While S-convLSTM reached the closest $K_i$/$V_b$ alignment, the B-convLSTM model achieved the lowest Patlak fitting error with less soft tissue shape distortion and produced $K_i$ estimates that resulted in the greatest ability to distinguish tumor malignancy. Moreover, With the smaller input and output feature map size, the B-convLSTM model is more practical in actual implementation with less GPU memory exhaustion. 

Since the motion-free gold standard dynamic images are not available for the real dataset, the motion impact on Patlak images was evaluated by running a significant motion simulation on a tumor phantom with pseudo local shift and expansion/contraction. As shown in the supplemental material, motion introduces overestimation of $K_i$. Thus, motion correction is expected to generate lower $K_i$ values in the lesion, which is consistent with our human study results presented in this paper.  

There are several limitations in this work, leading to further investigations. First, our current work focuses on the inter-frame motion that causes inter-frame mismatch, including the long-term respiratory and cardiac motion pattern change and the voluntary body movement. However, the intra-frame respiratory, cardiac, and body motion could also degrade the reconstructed image quality and lead to blur, introducing quantification error in parametric imaging as well. A potential future direction is combining the inter-frame and intra-frame motion correction, which is expected to further reduce the parametric fitting error. Given that the time and computation consumption of the current intra-frame motion correction method is very high \citep{mohy2015intra}, it will be worth investigating a deep learning approach and building a joint intra-inter-frame motion correction network. Second, the current proposed network only considers the inter-frame motion correction problem, and the target parametric estimation is then separately implemented by Patlak fitting. To fully take advantage of the tracer kinetics and the inter-frame temporal dependence, future work could include joint parametric imaging with the inter-frame motion correction to directly optimize the Patlak fitting error. Such an end-to-end framework would be more efficient and convenient for potential clinical applications. Lastly, all the current evaluations are based on a real dataset where there is no motion-free ground truth. By running motion simulations on generated motion-free phantoms, the motion correction performance would be more quantitatively evaluated.

In addition to the current limitations, some potential future directions are worth investigating. With other datasets under different tracers/imaging protocols, we could test the robustness of the proposed motion correction model. Transfer learning could also be applied for the model to generalize across different tracers, scanners, and imaging protocols. Also, the attention mechanism \citep{vaswani2017attention} has been proposed in natural language processing and shown its superiority in video capturing \citep{cai2020remote} and CT image segmentation \citep{oktay2018attention}. Introducing the attention mechanism into our motion correction model may help further correct the subareas with significant motion and improve the model performance. Finally, the backbone of the network is interchangeable and not limited to the U-Net used in this work. For example, Transformer based models have been widely used in a variety of natural language tasks \citep{gillioz2020overview}, and has been recently reported in image generation \citep{parmar2018image} and image restoration \citep{wang2021uformer}. It is worth investigating various backbone structures and further improve the model performance and efficiency.

\section{Conclusion}
In this work, we presented a new deep learning inter-frame motion correction method for dynamic PET and parametric imaging. Our proposed model integrated a convolutional LSTM layer into the bottleneck of a U-net style convolutional network to estimate motion across multiple dynamic PET frames. Compared with both traditional and deep learning baseline methods, the proposed convolutional LSTM model successfully further reduced the Patlak NFE and enhanced spatial alignment of the Patlak $K_i$ and $V_b$ images. In hypermetabolic ROIs, notable additional $K_i$ value changes and tumor texture changes were observed, with improved benign/malignant classification capability after correcting for motion with our proposed model. The well-trained network could perform motion estimation at around 7.4 seconds per frame, and the entire motion correction pipeline consumed only 1/12 of the wall time for the fully parallelized traditional BIS method and 1/170 of the overall compute time for BIS. Our proposed methods demonstrate the potential of introducing multiple-frame analysis and extracting spatial and temporal information for correcting motion in dynamic PET images.

\section*{Acknowledgments}
This work was under the support of the National Institutes of Health (NIH) through grant R01 CA224140 and a research contract from Siemens Medical Solutions USA, Inc. 

Bruce Spottiswoode, David Pigg and Michael E. Casey are employees of Siemens Medical Solutions USA, Inc. No other potential conflicts of interest relevant to this article exist.

\bibliographystyle{model2-names.bst}
\biboptions{authoryear}
\bibliography{refs}

\clearpage
\newpage
\renewcommand\thefigure{S\arabic{figure}} 
\renewcommand\thetable{S\arabic{table}} 
\section*{Supplementary Material}

The significant motion simulation was run on a synthetic tumor phantom to assess the motion impact on Patlak images. The pseudo local shift and expansion/contraction were implemented for simulating both rigid and non-rigid motion with random displacement magnitudes from -100mm to 100mm. To observe the motion impact more clearly, the maximum simulated motion magnitude was larger than the real dataset. We simulated one motion-free scan and five motion simulation scans, each acquiring 19 dynamic frames. The Patlak fitting was then implemented for the motion free and five motion simulation dynamic frame series. In Supplementary Table \ref{SupTable1}, the $K_i$ statistics and the NFE results were summarized in comparison between the motion simulations and the motion-free ground truth. In Supplementary Figure \ref{SupFig2}, the $K_i$ image and the NFE map of a sample motion simulation and the motion-free ground truth were visualized. The sample profile of the tumor was also visualized to directly display the motion impact on Patlak $K_i$ image.

\begin{table}[!h]
\centering
\setcounter{table}{0} 
\caption{Patlak $K_i$ statistics and NFE results in motion simulation (mean ± standard deviation) }
\label{SupTable1}

\begin{tabular}{@{}lcc@{}}
\toprule
 & Motion simulation & Motion-free ground truth\\ \midrule
$K_i$ mean & 0.0193 ± 0.0034 & 0.0146\\
$K_i$ maximum & 0.0514 ± 0.0073 & 0.0385\\
NFE & 8.5589 ± 0.9460 & 0.0187 \\
\bottomrule
\end{tabular}
\end{table}

\begin{figure}[!h]
\centering
\setcounter{figure}{0} 
\includegraphics[scale=0.45]{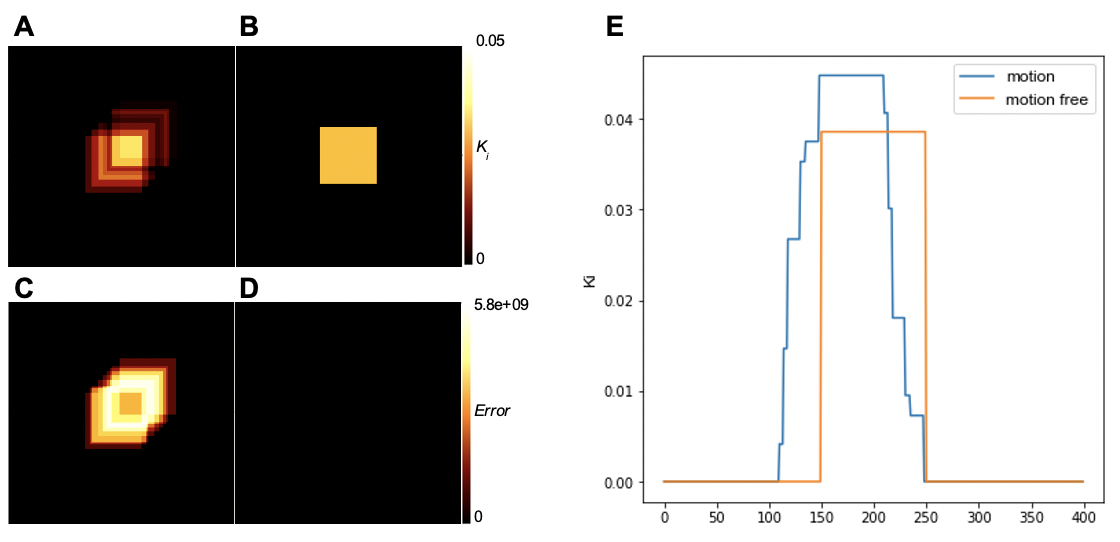}
\caption{Motion simulation results on a tumor phantom. (A) Sample Patlak $K_i$ image with simulated motion; (B) Sample motion-free Patlak $K_i$ image; (C) Sample NFE map with motion; (D) Sample motion-free NFE map; (E) Sample profile of the tumor at the middle of the slice, with motion simulation and motion free.}
\label{SupFig2}
\end{figure}

\begin{figure}[!t]
\centering
\includegraphics[scale=0.3]{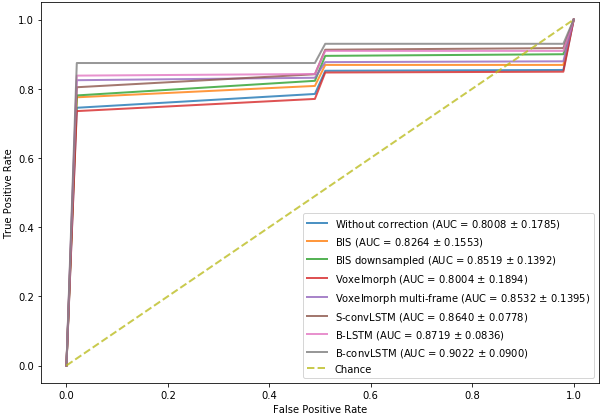}
\caption{The average ROC curve of each inter-frame motion correction method compared with chance. Note that due to  the small number of test samples in each of the 5 folds of stratified cross-validation, the number of distinct thresholds at which a new point on the ROC is generated is also relatively low.}
\label{SupFig1}
\end{figure}

\end{document}